\documentclass[10pt]{article}
\usepackage{amsmath}
\usepackage{upgreek}
\usepackage{cite}
\usepackage{times}
\usepackage{graphicx}

\begin{document}

\title{Analysis of damping-induced phase flips of plasmonic nanowire 
modes\footnote{\copyright 2012 OSA. This article was published in Optics Letters 
{\bf 37}(4), 746 (2012) and is made 
available with the permission of OSA. The 
paper can be found at the following URL on the OSA 
website: www.opticsinfobase.org/ol/abstract.cfm?uri=ol-37-4-746. 
Systematic or multiple reproduction or distribution to multiple 
locations via electronic or other means is prohibited and is subject to 
penalties under law.}}

\author{
Andreas Hohenau$^{1}$, Primoz Kusar, Christian Gruber, and Joachim R. Krenn \\
\small{Institute of Physics, Karl-Franzens-University, Universit{\"a}tsplatz 5, 8010 Graz, 
Austria} \\
\small{$^1$Corresponding author: andreas.hohenau@uni-graz.at}
}
\date{}

\maketitle

\begin{abstract}
We launch surface plasmons from one end of a silver nanowire by
asymmetric illumination with white light and use plasmon-to-light
scattering at the nanowire ends to probe spectroscopically the plasmonic
Fabry-Perot wire modes. The spectral positions of the
maxima and minima in the scattered intensity from both nanowire ends 
are found to be either in phase or out-of phase, depending on the 
nanowire length and the spectral range. This behavior can be explained by a generalized
Fabry-Perot model. The turnover-point between the two regimes 
is sensitive to the surface plasmon round trip losses and thus opens 
a new possibility for detecting changes of the optical absorption in the nanowire environment.
\end{abstract}

Surface plasmons (SP) are coherent oscillations of the electron density 
at the interface of a metal to a dielectric. In analogy to light in
dielectric fibers, SPs in the optical  spectral range
are guided by noble metal nanowires \cite{Dickson-Lyon:2000,Dorfm:NanoLett:2010}. In
contrast to dielectric structures plasmonic nanowires show however no
cutoff behavior as the SP wavelength scales with the nanowire
cross-section dimension
\cite{Takahara:OptLett:97}. Due to the confined SP electromagnetic fields, nanowires couple
efficiently to nearby photon emitters
\cite{Lukin:PRL2006,Akimov:2007,Kolesov:2009}, making them promising
building blocks in quantum optics. The coupling of light to nanowire SPs
(and vice versa) and the SP end-face reflection is governed by the geometry of the wire end faces
\cite{Li:NanoLett2009} and gives rise to SP Fabry-Perot (FP) modes, as has been observed by 
optical dark-field micro-spectroscopy\cite{Ditlbacher:PRL2005,Shegai:2011}. Here, we use this technique 
to %probe
%the FP resonances of SPs launched from one end of a silver nanowire. We
reveal a distinct phase-flip in the resonance positions measured from
both nanowire ends, whose spectral position is sensitive to the guided
SP mode round-trip losses.

% which allows to measure SP damping with high precision(?). The experimental observations are corroborated by a model simulation. 

\begin{figure}[htpb]
\center{\includegraphics[width=7cm]{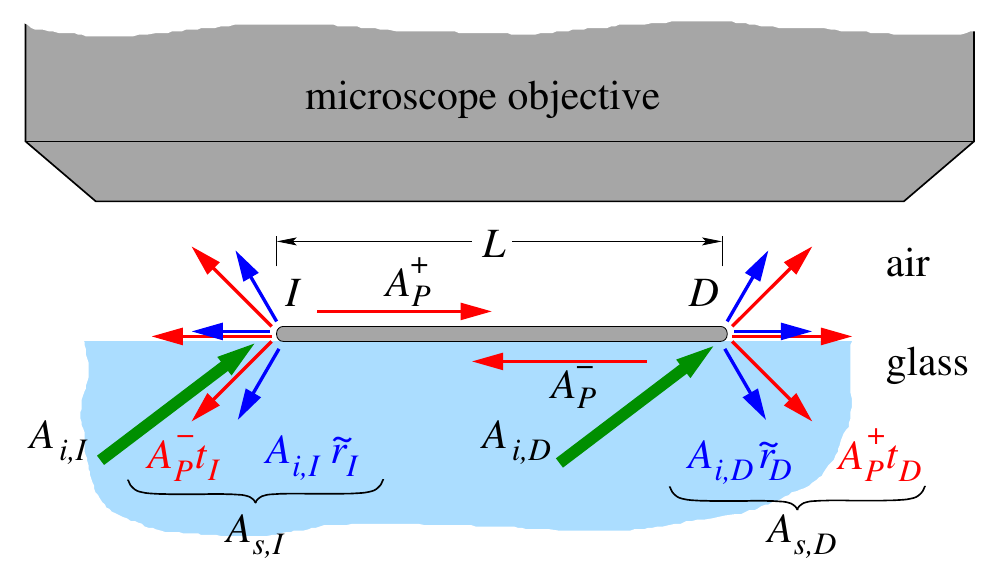}}
\caption{Sketch of a silver nanowire on glass substrate. The nanowire
was illuminated from the glass substrate in total internal reflection
geometry (thick, green arrows). The light scattered from the input ($I$) and
distal ($D$) end of the wire is collected by a microscope objective and
imaged on the input of a spectrograph. $A_{i,n}$ and $A_{s,n}$ are the
amplitudes of the light incident and scattered from the input ($n=I$) or
distal ($n=D$) end. $A_P^+$ and $A_P^-$ are the amplitudes of the
excited right and left propagating SP mode, respectively, and $r$ and
$t$ are the reflection, transmission and scattering efficiencies as
described in the text.}
\label{fig:sketch}
\end{figure}

Silver nanowires were chemically synthesized by a seed-assisted
reduction of AgNO$_3$ with ethylene glycol in the presence of poly(vinyl
pyrrolidone) \cite{Dapeng:JMS2011}. The resulting wires with
cross-section dimensions around 100 nm and length up to a few tens of
$\upmu$m were spin cast in an ethanol solution on glass substrates. The
substrates were optically immersed to a glass prism and illuminated with
a slightly focused white light beam from a halogen lamp under total
internal reflection. This excitation geometry %ensures that SPs are
%launched only from the \textit{input} ($I$) nanowire 
%end~\cite{Ditlbacher:PRL2005}, as defined in
%Fig.~\ref{fig:sketch}. In addition, it 
allows a dark field detection, as
the microscope objective (100x, numerical aperture 0.95) only collects
light individually scattered from the {\it input} ($I$) and {\it distal} ($D$)
nanowire ends (Fig.~\ref{fig:sketch}). The collected light is dispersed
by a spectrophotometer and detected by a camera. For all experiments the
excitation light is polarized in the plane of incidence which coincides
with the long nanowire axis.

\begin{figure}[htpb]
\center{\includegraphics{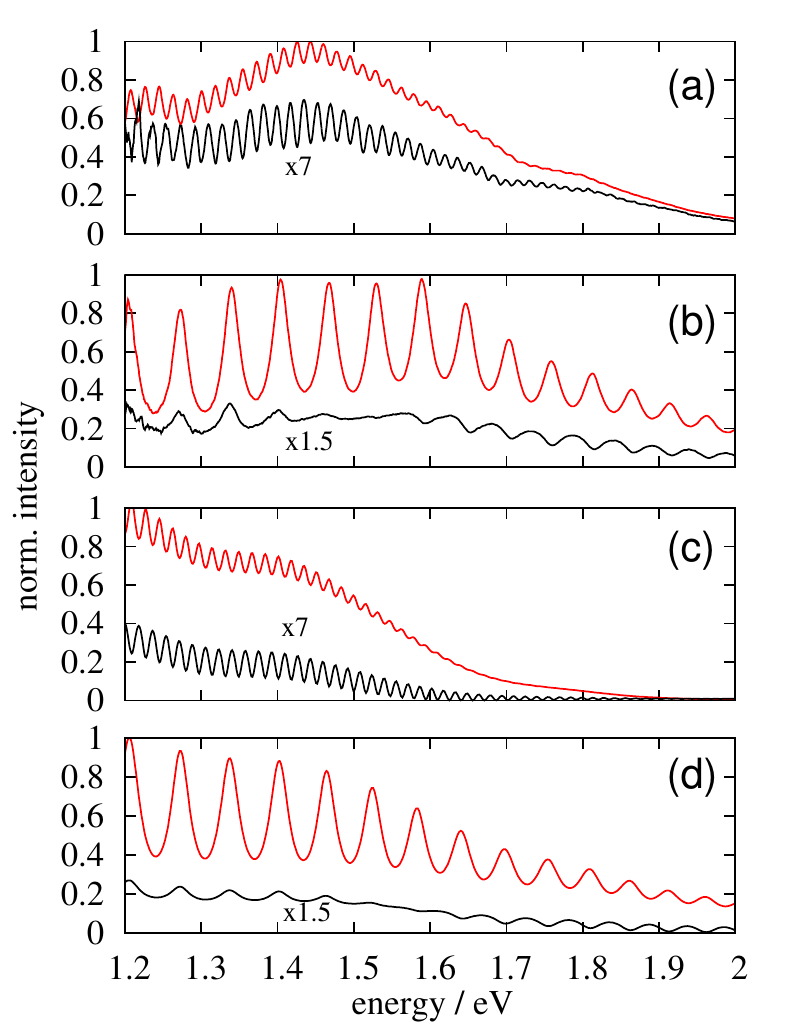}}
\caption{(a,b): Experimentally recorded scattering spectra of (a) a
$19\:{\rm \upmu m}$ and (b) a $5\:{\rm \upmu m}$ long silver nanowire of the
input (lower, black curves) and distal (upper, red curves) ends. (c,d): Theoretical
scattering spectra calculated by a generalized FP model for (c) a
$19\:{\rm \upmu m}$ and (d) a $5\:{\rm \upmu m}$ long nanowire (see text).}
\label{fig:spectra} 
\end{figure}

Fig.~\ref{fig:spectra} depicts the experimentally recorded scattering
spectra for two nanowires of $19\:{\rm \upmu m}$ (a) and $5\:{\rm \upmu m}$ (b) length.
The intensities recorded from the distal wire ends $I_{s,D}$ show the
typical periodic modulations attributed to SP-FP 
resonances~\cite{Ditlbacher:PRL2005}. For
the $19\:{\rm \upmu m}$ long nanowire the (weaker) intensity scattered from the
input end ($I_{s,I}$) is in counter-phase to $I_{s,D}$ scattered at the 
distal end. This is
generally expected in analogy to a FP-interferometer from the 
viewpoint of energy conservation. If the transmission (corresponding
to $I_{s,D}$  in the nanowire case) is large, the reflection (corresponding
to $I_{s,I}$  of the nanowire case) has to be small and vice
versa~\cite{Ditlbacher:PRL2005}. However, for the $5\:{\rm \upmu m}$ long nanowire,
$I_{s,I}$ changes from being in-phase with $I_{s,D}$ at photon energies
below  $\sim 1.55\:{\rm eV}$ to being in counter-phase above which
deviates from the expectations for a FP-interferometer.

To gain a qualitative understanding of the observed behavior, we develop
in the following a generalized FP model suitable for describing the
scattering at the nanowire ends. In analogy to a FP-interferometer,
the following (in general frequency dependent) coefficients are defined (Fig.~\ref{fig:sketch}):
\begin{itemize}
\item{The direct scattering efficiency $\tilde{r}_n=A_{s,n}/A_{i,n}$ for the
in-coupling ($n=I$) and distal ($n=D$) nanowire ends, where $A_{s,n}$ is
the (complex valued) amplitude of the locally  scattered light and
$A_{i,n}$ is the amplitude of the excitation light at the respective nanowire 
end.} 
\item{The SP mode excitation efficiency
$\tilde{t}_n=A_{P,t,n}/A_{i,n}$ at the nanowire ends, where
$A_{P,t,n}$ is the amplitude of the SP mode excited at end $n$.}
\item{The SP mode reflection coefficient at the nanowire ends
$r_n=A_{P,r,n}/A_{P,i,n}$, where $A_{P,r,n}$ and $A_{P,i,n}$ are the
amplitudes of the reflected and incident SP at the nanowire end $n$, 
respectively.} 
\item{The SP scattering
efficiency $t_n=A_{s,P,n}/A_{P,n}$, with $A_{s,P,n}$ being the amplitude of the
light emitted from the nanowire ends due to scattering of the incident SP mode with amplitude $A_{P,n}$.} 
\end{itemize}

With $A_P^+$ and $A_P^-$ denoting the amplitudes of the left- and right-propagating SP 
modes, one obtains: 
%If the amplitudes of the left- and right-propagating SP modes are 
%written as $A_P^+$ and $A_P^-$, one gets the following relations:
\begin{equation}
\begin{array}{rcl}
A_{i,I} \tilde{t}_I+A_P^- r_I & = & A_P^+ \\
A_{i,D} \tilde{t}_D+A_P^+e^{ikL} r_D & = & A_P^-e^{-ikL}\\
A_{s,D} & = & A_P^+ e^{ikL} t_D + A_{i,D} \tilde{r}_D \\
A_{s,I} & = & A_P^- t_I + A_{i,I} \tilde{r}_I
\end{array}
\end{equation}
This leads to 
\begin{equation}
\begin{array}{l}
A_{s,I}  =  A_{i,I} \tilde{r}_I +e^{ikL}\frac{A_{i,D}\tilde{t}_D t_I 
+A_{i,I}r_D t_I\tilde{t}_Ie^{ikL}}{1-r_I r_D e^{2ikL}} 
\\
A_{s,D}  =  A_{i,D} \tilde{r}_D +e^{ikL}\frac{A_{i,I}\tilde{t}_I t_D 
+A_{i,D}r_I t_D\tilde{t}_De^{2ikL}}{1-r_I r_D e^{2ikL}} 
\end{array}
\end{equation}
For the following, we make the simplification $r_I=r_D=r$, $t_I=t_D=t$ (justified for 
identical shapes of both nanowire ends) and the assumption that $\tilde{t}_D \simeq 
0$~\cite{Ditlbacher:PRL2005} and $\tilde{r}_D \simeq 0$. Already small 
deviations from this assumption cause alternating larger and smaller 
amplitudes of neighboring peaks in the scattering spectra of $I_{s,I}$ 
and/or $I_{s,D}$, which is not observed in Fig.~\ref{fig:spectra}.
Consequently, the related intensities are 
\begin{equation}
\begin{array}{rcl}
I_{s,D} & \propto & |A_{s,D}|^2 \propto I_i \left|\frac{\tilde{t}_I t}{1-r^2e^{2ikL}}\right|^2  %{1-2Re^{-2k''L}\cos(2k'L+2\psi)+R^2e^{-4k''L}}.
 \mbox{ and}   \\ 
I_{s,I} & \propto  & |A_{s,I}|^2 \propto I_i |\tilde{r}_I|^2 
\left|\frac{1-(r^2-rt\tilde{t}_I/\tilde{r}_I)e^{2ikL}}{1-r^2e^{2ikL}}\right|^2 

\end{array}
\label{eq:Ints}
\end{equation}
%For the latter relation $\left| \frac{1}{1-Z} 
%\right|^2=\frac{1}{1-2Z'+|Z|^2}$ with $Z=Z'+iZ''=r^2e^{2ikL}$ and $r^2=Re^{2i\psi}$ was used.
%The $\omega^4$ dependence of dipole radiation is not considered.

We note that for an actual experimental setup the light collection
efficiency given by the numerical aperture of the detection optics and
the scattering diagram of $I_{s,n}$ ($n=I,D$) can be included to the
above equations as additional factors to $\tilde{r}_n$ and $t_n$.

For comparison to the experimental results, we plot in
Fig.~\ref{fig:spectra}(c,d) the scattering spectra as calculated by
Eq.~\ref{eq:Ints}. The complex SP-mode dispersion relation was 
approximated by that for a $100\:{\rm  nm}$ diameter Ag 
wire~\cite{Takahara:OptLett:97,Palik:Book} in a homogeneous dielectric 
environment. To roughly reproduce the experimentally observed spectral 
periodicity, we used $1.85$ as effective dielectric constant. The
resulting SP-mode propagation  length was additionally multiplied by a
factor of 1.7 to better resemble experimental
values~\cite{Kusar:NanoLett:2011}. Further, the $r$ and $t$ coefficients
are assumed purely real and frequency independent as
$r=\sqrt{0.4}$~\cite{Kusar:NanoLett:2011} and $t=\sqrt{1-r^2}$ (energy
conservation at the end-cap). $\tilde{r}_I$ was set to
$-0.04\tilde{t}_I$  ($-0.01\tilde{t}_I$) for the $5\:{\rm \upmu m}$
($19\:{\rm \upmu m}$). The different but comparable small values of
$\tilde{r}_I$ were chosen for  a better accordance with the experiment and
might be indicative for different end cap geometries of the two wires. Despite the
simplified  assumption of frequency independent scattering and
reflection  coefficients (responsible for the discrepancies in 
``background'', modulation depth and the exact peak positions), good
qualitative correspondence is observed between theory and experiment.
This indicates, that the developed model indeed describes well the
physics of the nanowire SP-FP-resonators. 

As described above, particularly for the $5\:{\rm \upmu m}$ long nanowire
$I_{s,I}$ changes its qualitative behavior from being in-phase with
$I_{s,D}$ at photon energies below  $\sim 1.8\:{\rm eV}$ to being in
counter-phase above in both theory and experiment. This behavior is not
expected from a FP-interferometer but is a peculiarity of the nanowire
due to two reasons. First, the efficiency of direct scattering from the
nanowire input end $\tilde{r}_I$ (analog to the external reflectivity of a
FP-interferometer) is much smaller than the SP reflection $r$,
whereas in a usual FP-interferometer they are identical. Second, the
excited SP modes suffer from propagation losses which are usually
neglectable in a FP-interferometer. In combination of both, for a short
nanowire (e.g. the $5\:{\rm \upmu m}$ long wire) and in the red spectral
range the SP round trip losses are small and the amplitude of
the externally scattered light is neglectable against the
amplitude of the light coupled out from the SP-FP mode. Thus the
intensity of light scattered from both nanowire ends is dominated by the
SP intensity only and maxima appear at the same spectral positions for
both input and distal end. In contrast, for larger nanowire length (e.g.
the $19\:{\rm \upmu m}$ long wire) or in the blue spectral range of the 
spectrum for the 
$5\:{\rm \upmu m}$ long wire, the amplitude of the SP-mode and thus that
of the outcoupled light gets, due to ohmic losses, comparable or even
smaller than the amplitude of the directly scattered light at the input 
end. If, as in
a FP-interferometer, both are in counter-phase they interfere
destructively. In the spectrum this leads to minima of the scattered
light intensity at the input end at those spectral positions where the
scattered light intensity of the distal end has maxima.

\begin{figure}
\center{\includegraphics{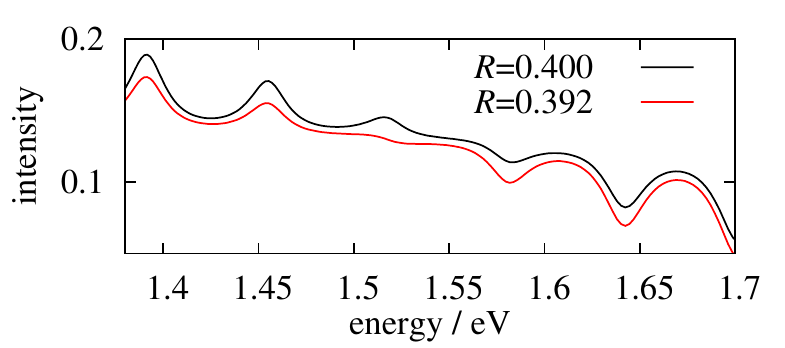}}
\caption{Calculated spectra of 
$I_{s,I}$ around the phase flip point for
a $5\:{\rm \upmu m}$ long nanowire. The SP end cap reflectivity is assumed
as $R=|r|^2=0.4$ (upper, black curve) and $R=0.392$ (lower, red curve). The resulting
spectral shift of the phase flip point leads to a pronounced qualitative
difference between the curves around this point.}
\label{fig:sensing}
\end{figure}

For the $5\:{\rm \upmu m}$ long nanowire, both domains are within the
spectral range, and the intensity profiles of $I_{s,I}$ changes its
behavior from being in-phase with $I_{s,D}$ below $\sim 1.55\:{\rm eV}$
to being in counter-phase above. The condition for this phase-flip
can be derived from Eq.~\ref{eq:Ints} as~\cite{footnote} 
\begin{equation}
\tilde{r}_I=-\frac{t \tilde{t}_I}{r} \frac{\xi^2}{1-\xi^2}
\label{eq:I_I}
\end{equation}
or
\begin{equation}
\xi^2=\frac{\tilde{r}_I r}{\tilde{t}_I t + \tilde{r}_I r}
\end{equation}
with $\xi=R|e^{2ikL}|$ as the SP-amplitude round trip losses. At this point 
one finds 
\begin{equation}
 I_{s,I}=I_i|\tilde{r}_I|^2 \frac{1}{\xi^2},
 \label{eq:isi}
\end{equation}
and, for the ratio of the amplitude of the light scattered from the SP 
mode at the input end to that of the directly scattered light
\begin{equation}
 \left| \frac{A_P^- r}{A_{i,I} \tilde{r}_I} \right |=1+\frac{1}{\xi}.
\end{equation}

In contrast to the spacing of the maxima and minima in $I_{s,D}$ which
depend on the round-trip phase, the spectral position of the phase-flip
in $I_{s,I}$ depends on the SP round-trip losses only. For $\sim 2-7\:{\rm \upmu m}$ long 
wires it could be used
(in combination with the round trip losses determined from the spectra 
taken from the distal end~\cite{Kusar:NanoLett:2011}) to deduce the modulus of 
the direct scattering
efficiency $|\tilde{r}_I|$  by Eq.~\ref{eq:isi}, if the scattered 
intensity at the input end is measured quantitatively. Another 
application would be in nanowire sensor concepts as an information
channel on the nanowire SP-mode propagation- and reflection losses to
assess changes in the absorption of the nanowire environment or in the
end cap reflectivity (for example caused by adsorbates or changes in the
cap's geometry). Fig.~\ref{fig:sensing} depicts the results of our
calculations of $I_{s,I}$ around the phase-flip point, where the end-cap
reflection $r$ was lowered by $1\%$ (red curve, the round-trip losses 
$\xi$ are thus increased by $4\%$) leading to a pronounced qualitative
change in $I_{s,I}$.

In conclusion, we demonstrated a phase change of the intensity scattered
at the input end of a several ${\rm \upmu m}$ long  Ag nanowire compared
to that scattered at the output end. By developing a generalized FP model
adapted to the metal nanowire case, this feature can be explained by a
combination of propagation losses of the excited SP modes and a
relatively low efficiency for direct (i.e. without the
involvement of the SP-FP mode) scattering of the exciting light at the input end.
The spectral position of the phase-flip point is sensitive to the
SP-mode round trip losses in the metal nanowire. It thus could be
applied in  nanowire sensors to assess changes in the SP losses in
addition to the SP propagation constant probed by the spectral spacing
of the transmission maxima. 

\section*{Acknowledgement}
Financial support from the European Union project ICT-FET 243421 ARTIST 
and from the Austrian Science Fund (FWF): P21235-N20 is acknowledged.
%\pagebreak
%\bibliography{krenn,kusar,hohenau}

\begin{thebibliography}{10}
\bibitem{footnote}
Around this point 
$I_{s,i}$ %\propto |(1-A)/(1-B)|^2=|1-A|^2/|1-B|^2$ 
has to be independent of $k$. %(with $A=(r^2-rt\tilde{t}_I/\tilde{r}_I)e^{2ikL}$, $B=r^2e^{2ikL}$. 
%With the relations $|1-A|^2=1-2A'+|A|^2$ and
%$|1-B|^2=1-2B'+|B|^2$, one obtains ${2A'}/{2B'}=({1+|A|^2})/({1+|B|^2})$.
By equating the coefficients of the cosine functions with identical
periodicity in the denominator and numerator of $I_{s,i}$, 
%$|1-A|$ and $|1-B|$ and by introducing the abbreviation
%$(1-c)e^{i\phi}=(1-t\tilde{t}_I/r\tilde{r}_I)$
one obtains after some algebra %$c=1-1/\xi^2$ (with $\phi=0$ and $\xi=R|e^{-2ikL}|$) and finally
Eq.~\ref{eq:I_I}.
\end{thebibliography}

\end{document}